\documentstyle[aps,prl,twocolumn,epsfig]{revtex}
\newcommand{\beq}{\begin{equation}}
\newcommand{\eeq}{\end{equation}}
\newcommand{\bea}{\begin{eqnarray}}
\newcommand{\eea}{\end{eqnarray}}
\newcommand{\beas}{\begin{eqnarray*}}
\newcommand{\eeas}{\end{eqnarray*}}

\begin{document}
\draft
\wideabs{
\title{Anderson impurity in a correlated conduction band}
\author{W.~Hofstetter, R.~Bulla, and D.~Vollhardt}
\address{Theoretische Physik III, Elektronische Korrelationen und
Magnetismus, Universit\"at Augsburg, 86135 Augsburg, Germany}
\date{\today}
\maketitle
\begin{abstract}
We investigate the physics of a magnetic impurity with spin $1/2$ in
a correlated metallic host. Describing the band by a Hubbard Hamiltonian, 
the problem is analyzed using dynamical
mean-field-theory in combination with Wilson's nonperturbative 
numerical renormalization group. We present results for the 
single-particle density of states and the dynamical 
spin susceptibility at zero temperature. New 
spectral features (side peaks) are found which should be observable 
experimentally. In addition, we find a general enhancement 
of the Kondo scale due to correlations. Nevertheless, 
in the metallic phase, the Kondo scale always vanishes 
exponentially in the limit of small hybridization.
\end{abstract}

\pacs{PACS numbers: 71.27.+a, 75.20.Hr}

}
The Anderson model \cite{Anderson 61} has been successfully applied in 
the past to describe 
the physics of a magnetic impurity embedded in a conducting host. 
Extensive theoretical studies of this particular many-body problem 
led to considerable insight as well as progress in the development of new
methods\cite{Hewson 93}. The most thoroughly analyzed case 
is that of an impurity in a noninteracting conduction band with a constant density 
of states. Several properties have been established by Wilson's numerical 
renormalization group \cite{Wilson 75,Krishnamurthy 80} and by the Bethe ansatz 
\cite{Wiegmann 81}. Most importantly, a new many-body energy scale
$T_K$ (the Kondo temperature) arises, 
which is exponentially small in the
limit of vanishing hybridization. In addition, it has been demonstrated
\cite{Nozieres 74} that below this temperature the system can always be 
understood as a ``local Fermi liquid'' with strongly renormalized quasiparticles. 
The single-particle spectrum was shown to exhibit a generic 
three peak structure consisting of two atomic levels and a quasiparticle
resonance of width $T_K$. 
Based on this model, a number of experimental results for dilute
impurities in metals have been explained successfully, including 
measurements of the resistivity, the magnetic susceptibility and 
the specific heat. For a review see ref.~\cite{Tsvelick 83}.

It has become clear, however, that the single impurity Anderson model
is somewhat too simplified and that qualitatively different 
types of physical behaviour 
are possible when a more general Hamiltonian is considered. 
One very important characteristic of real materials is the interaction 
among the conduction electrons. 
This aspect is usually neglected, mostly for technical reasons,
i.e.~to simplify the investigation. If taken into account, 
we expect, at least, a renormalization of the model parameters. 
Our work will focus on the question whether, in addition, 
qualitatively new physics is possible. 
An experimental realization frequently cited in this context 
is the cuprate system 
$\textrm{Nd}_{2-x}\textrm{Ce}_x \textrm{Cu} \textrm{O}_4$ \cite{Brugger 93}, 
a concentrated 
impurity system, where the energy scale of low temperature 
heavy fermion behaviour is apparently incompatible with the 
standard Kondo picture.

Models with a single impurity embedded in a correlated host were 
studied already within several approaches. 
Perturbative calculations in a slave boson representation 
by Khaliullin and Fulde\cite{Khaliullin 95} yielded a 
renormalization of the effective Kondo coupling. 
Very similar results were obtained by Tornow at al.~\cite{Tornow 97}
within a non-crossing approximation.
Furthermore, in the limit of high dimensions and using a variational
treatment, Davidovich and Zevin\cite{Davidovich 98} found 
a qualitative change of the behaviour of the Kondo
temperature $T_K$.
According to these authors, above some intermediate value of the 
conduction band interaction, $T_K$ is no longer exponentially small 
at vanishing exchange coupling. 
In our work, we will discuss this issue in detail. 
Finally, in the case of one dimension, Phillips and Sandler
\cite{Phillips 96} and also Schiller and Ingersent \cite{Schiller 95}
represented the interacting host as a Luttinger liquid, which makes a 
renormalization group treatment possible. 
Among other results, they found that in some region of 
parameter space an unquenched local moment may survive down 
to the lowest temperatures.
\begin{figure}[t]
\begin{minipage}{0.47\linewidth}
\begin{center}
\epsfig{file=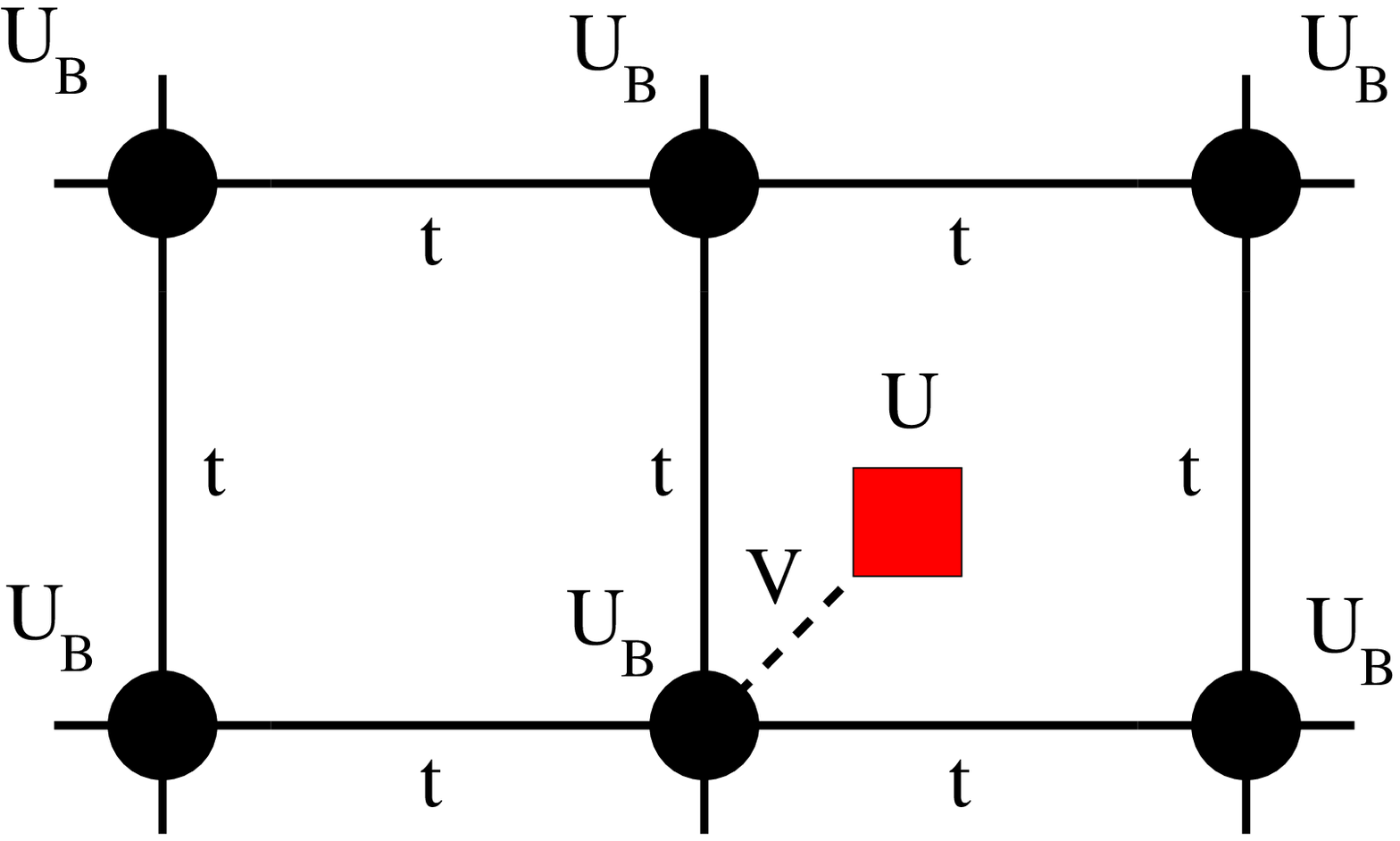,width=\linewidth}
\end{center}
\end{minipage}
\hfill
\begin{minipage}{0.4\linewidth}
\begin{center}
\epsfig{file=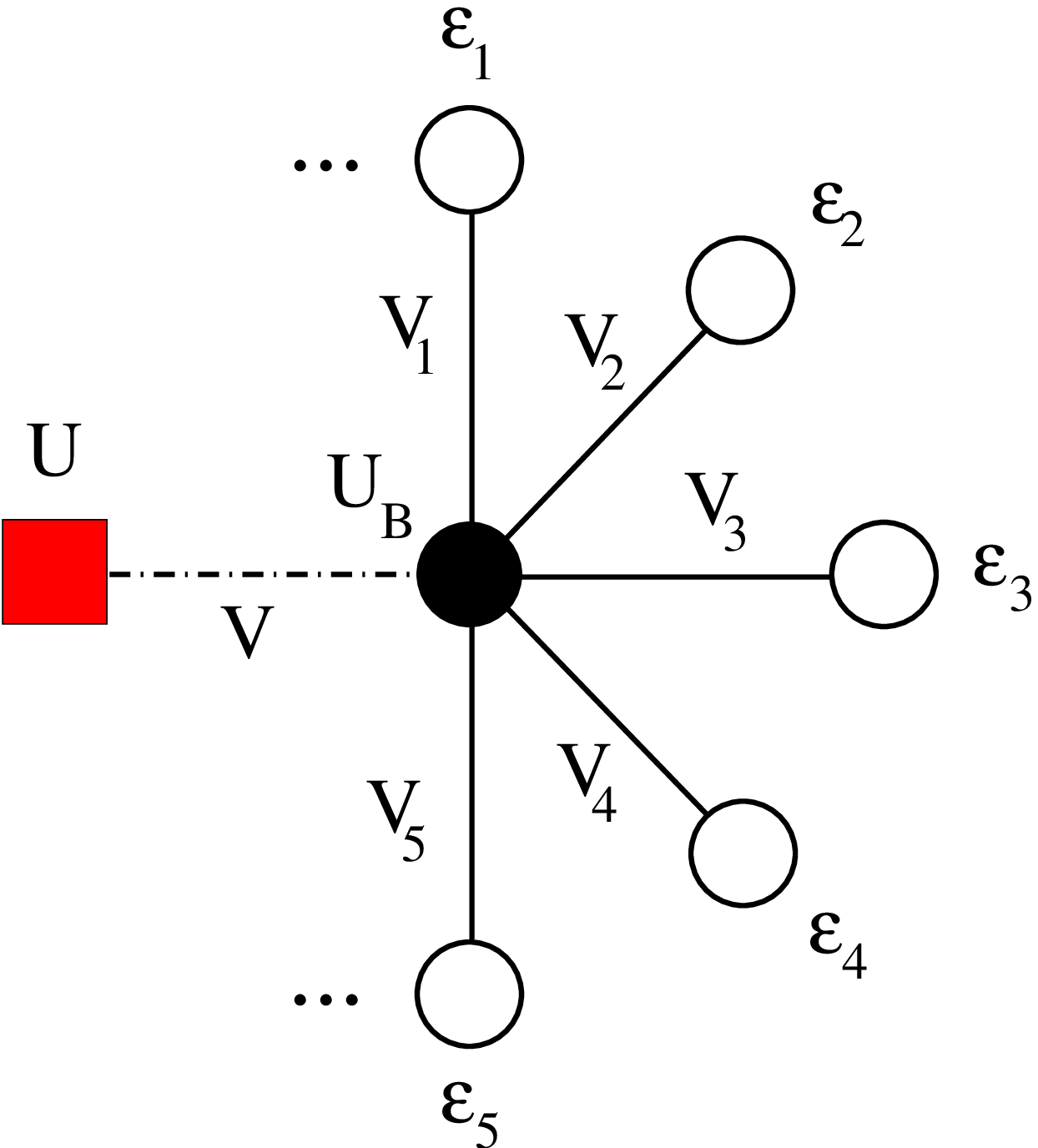,width=\linewidth}
\end{center}
\end{minipage}
\caption{\label{fig:graphik}Left: Anderson impurity (square) coupled
to one site of a correlated lattice (circles). The hybridization is
taken to be purely local. Right: effective two-impurity model with noninteracting band orbitals 
(empty circles) of energy $\epsilon_i$ and hybridization $V_i$.
} 
\end{figure}
These studies already indicate the competition of 
several effects: (i) The conduction band correlations may 
change the density of states (DOS) of the conduction band. 
(ii) A repulsive on-site interaction will reduce the  
hybridization of the impurity level. 
(iii) The conduction electrons will 
become increasingly polarized, thus enhancing the 
effective spin coupling of the impurity moment. 
In the following, using Wilson's nonperturbative 
numerical renormalization group,
we will analyze which one of these 
factors dominates.

Our Hamiltonian consists of a spin-$1/2$ impurity embedded in an 
interacting host (see also fig.~\ref{fig:graphik}) represented by a one-band Hubbard 
model\cite{Gutzwiller 63}:
\bea \label{total_hamiltonian}
H &=&  -\sum_{i j \sigma} \left(t_{i j} - \epsilon_c\, \delta_{i j}\right)\, 
c^\dagger_{i \sigma}\, c^{\phantom{\dagger}}_{j \sigma} + 
U_B\, \sum_{i} n_{i\uparrow}\, n_{i\downarrow}  \nonumber \\
&&+ V\sum_\sigma \left(f_\sigma^\dagger\, c^{\phantom{\dagger}}_{0 \sigma} +
h.c.\right) + U\, n_{f\uparrow}\, n_{f\downarrow} + 
\epsilon_f\, n_f.
\eea
Note that the impurity hybridizes with a 
single conduction band orbital, which in the following will be denoted 
as the $i=0$ Hubbard site.

We will be interested in the case of half filling, 
which  -- assuming a bipartite lattice and next-neighbour hopping 
($t_{i j}=t$) only -- is equivalent to 
$\epsilon_c = -U_B / 2$ and \mbox{$\epsilon_f = -U/2$}.
The calculations in our paper are restricted to the paramagnetic 
phase of the host.
A controlled approach\cite{Metzner 89,Georges 96} to correlated lattice problems is possible in
the limit of large coordination number $Z\to\infty$, scaling the 
hopping matrix elements as $t = \frac{t^*}{\sqrt{Z}}$.  
In our treatment we will use the Bethe lattice and take the
noninteracting half-bandwidth $D=2 t^*=1$ as the unit of energy.
It should be emphasized, however, that the choice of the lattice 
is merely motivated by calculational convenience and should 
have no qualitative effect on the results\cite{Bulla 99b}.

It was pointed out in \cite{Davidovich 98} that 
integrating out all the band fermions except those on the 
$i=0$ Hubbard site (the so-called ``cavity method'') yields an effective action 
which has the same retarded part
${\cal S}_{\rm eff,ret} = -\int\!\!\!\int d\tau\, d\tau'\, 
c^\dagger_{0\sigma}\, {\cal G}_0^{-1}(\tau - \tau')\, 
c_{0\sigma}(\tau')
$ 
as the pure Hubbard model. 
As a result, the system can be described as a two-impurity model 
(see fig.~\ref{fig:graphik}) with an effective  
\emph{noninteracting} bath defined by a hybridization function
$
\Delta_c(\omega) = \pi \sum_p  |V_p|^2\, \delta(\omega - \epsilon_p).
$
In our calculation we will therefore follow a two-step procedure: 
First, we solve the Hubbard model in dimension $d\to\infty$ 
using Wilson's numerical renormalization group\cite{Hewson 93,Krishnamurthy
80} as in\cite{Bulla 99b}. 
In the paramagnetic phase considered here this leads to a Mott transition 
at a critical interaction strength $U_B^c \approx 2.92$. 
Correlations strongly influence the structure of the 
DOS; close to the transition point 
an effective narrow band\cite{Hofstetter 99}
is formed by the quasiparticle resonance.

In the next step, we add the $f$-impurity. 
The combined system is then again treated using NRG, 
this time without the self-consistency loop 
(the modification of the effective bath due to the 
impurity is ${\cal O}(1/N)$ and can therefore be neglected 
in the thermodynamic limit). The band 
correlations enter via the previously determined DOS and 
the $c$-site interaction $U_B$. In our calculations we have 
confined ourselves to the metallic regime $U_B < 2.92$ and 
zero temperature.

First, we present results for the single particle spectra 
$\rho_{f(c)}(\omega) = -\frac{1}{\pi}\, {\rm Im} G_{f(c)}(\omega)$ 
of the impurity $f$ and the Hubbard site $i=0$. 
Considering $\rho_f$ (fig.~2a), we obtain a three-peak 
structure in the spectrum already for \mbox{$U=0$} 
and small hybridization $\Delta = \frac{\pi V^2}{2 D}$. 
This may be attributed to a narrowing of the effective
band, leading to resonances at finite energy\cite{Hofstetter 99}. 
Upon increasing $U$, these peaks are reduced and weight is shifted to 
the atomic levels which for large interaction 
can be found at $\omega \approx \pm U/2$.
In addition, the height of the quasiparticle peak is significantly reduced. 
Luttinger's theorem\cite{Luttinger 61} which states 
that $\rho_f(0)$ is pinned at its $U=0$ value is therefore 
found to be not valid in the case of an interacting conduction band.

Furthermore, we notice that the width of the quasiparticle 
resonance is almost independent of $U$, in contrast to 
the situation with $U_B=0$. This already indicates a strong 
enhancement of the Kondo scale due to band correlations, 
as will be discussed in more detail below.
\begin{figure}[t]
\begin{center}
\epsfig{file=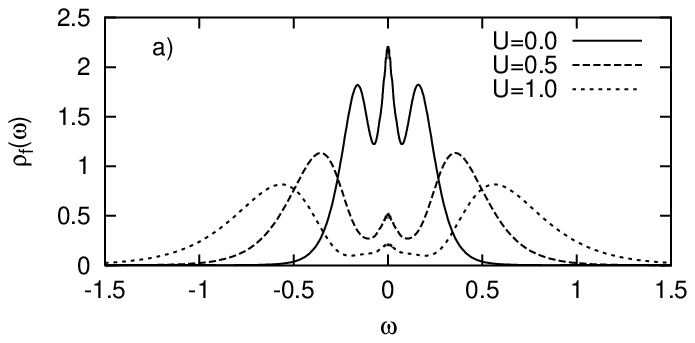,width=0.9\linewidth}
\epsfig{file=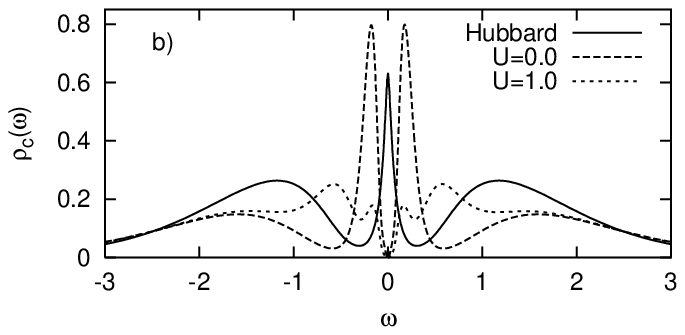,width=0.9\linewidth}
\end{center}
\caption{\label{fig:spectral_density}Spectral densities 
$\rho_f$ (a) and $\rho_c$ (b) at zero
temperature, $\Delta=0.1$ and $U_B=2.6$. 
For comparison, we also show the $c$ spectral density 
at vanishing hybridization $\Delta$, equivalent to the 
DOS of the Hubbard model.
}
\end{figure}
The corresponding results for $\rho_c$ are shown in 
fig.~2b. 
For any finite $\Delta$ a hybridization gap is formed. 
This is already the case at $U=0$ and $U_B=0$ (not shown here) 
and persists for finite interactions, 
indicating that the system is a Fermi liquid 
(a nonvanishing self-energy at $\omega=0$ would 
smear out the gap).
The Fermi liquid picture is independently supported by the fact 
that 
the fixed point of the NRG 
and its leading irrelevant eigenoperators are unchanged 
compared to the noninteracting case; see also \cite{Takayama 98}.
%For any finite $\Delta > 0$ 
%a hybridization gap is formed and the spectral density 
%vanishes exactly at the Fermi level. 
%This is consistent with a Fermi liquid behaviour of the self 
%energy, which is independently supported by the fact that 
%the fixed point of the NRG 
%and its leading irrelevant eigenoperators are unchanged 
%compared to $U_B=0$.

A quantity which is more easily accessible experimentally is 
the (longitudinal) dynamic susceptibility, defined as the response of 
the impurity spin to a local magnetic field
\beq
\chi(\omega) = i \int\limits_0^\infty dt\, e^{i \omega t}\, 
\langle \left[S_f^z(t),S_f^z(0)\right]\rangle.
\eeq
Within the NRG formalism it is convenient\cite{Costi 99b} to calculate the 
imaginary part $\chi''(\omega)$ directly and to obtain 
the real part $\chi'(\omega)$ via Kramers-Kronig transformation.
In particular, the static susceptibility is then given by 
$\chi_0 = \chi'(0)$.
In the following we will focus on the spin relaxation function
\beq
S(\omega) = \frac{\chi''(\omega)}{\pi \omega}.
\eeq
In a first step, we consider its behavior with increasing band 
correlations for fixed $\Delta, U$ (see fig.~3a). 
\begin{figure}[t]
\begin{center}
\epsfig{file=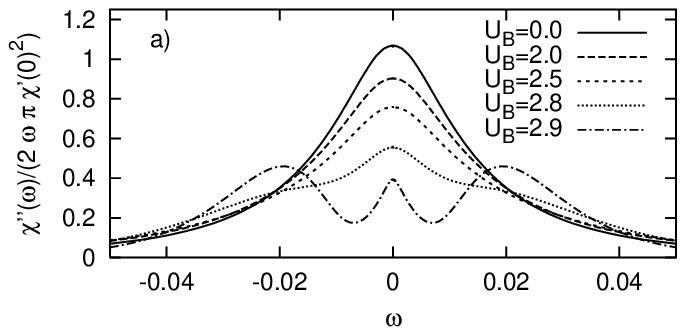,width=0.9\linewidth}
\epsfig{file=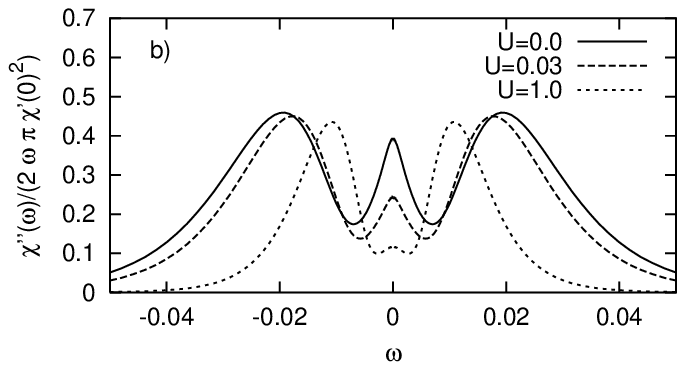,width=0.9\linewidth}
\end{center}
\caption{\label{fig:spinrelaxation}Local spin relaxation function 
for fixed $U=0$ (a) and $U_B=2.9$ (b). In both cases, the
hybridization has the value $\Delta=0.01$. Note that due to 
the normalization a value of $1$ at $\omega=0$ would be consistent 
with the Shiba relation.}
\end{figure}
In this plot we employed a normalization of $\chi''(\omega)$ 
suggested by the Shiba relation\cite{Shiba 75,Yoshimori 82} 
\beq
\lim_{\omega\to 0} \frac{\chi''(\omega)}{\pi \omega} = 
2\, \chi_0^2
\eeq
derived for $U_B=0$\cite{Yoshimori 82}.
For the noninteracting band this relation is 
indeed satisfied (with an error of less than 10\%
due to the NRG procedure). With increasing $U_B$ systematic deviations 
arise, indicating that the general proof\cite{Yoshimori 82}
based on Ward identities breaks down for an interacting conduction band.
The lineshape of $\chi''(\omega)$ also depends on $U_B$. 
For a weakly correlated band we obtain a single elastic peak, 
while close to the metal-insulator transition at $U_B^c=2.92$ two
additional inelastic side peaks arise. 
They indicate that in this case the Kondo singlet is formed at an 
energy scale which lies outside the effective band. 
The width of the remaining elastic peak 
is determined by the effective bandwidth of the Hubbard model.

%This can be seen clearly by comparing the above results to those
%for a noninteracting, but narrow, band (not shown here). 
An increase of the impurity interaction $U$ (fig.~3b), 
leads to a suppression of the elastic peak and to
a shift of the inelastic peaks
(corresponding to a slight reduction of the singlet binding energy).

\begin{figure}[t]
\begin{center}
\epsfig{file=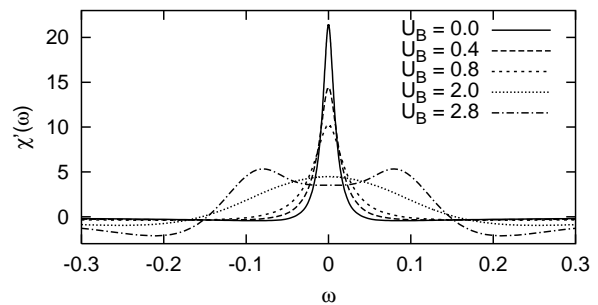,width=0.9\linewidth}
\end{center}
\caption{\label{fig:chi_1}Real part of the dynamic susceptibility
at $\Delta=0.1$ and $U=1.0$.}
\end{figure}

For the real part $\chi'(\omega)$, some typical 
results are shown in fig.~\ref{fig:chi_1}: Already at a weak 
band interaction $U_B$ (when the Hubbard DOS is well approximated 
by the noninteracting one) the static susceptibility
$\chi_0$ is strongly reduced.
Of particular interest is the hybridization dependence of the low energy scale, 
the Kondo temperature $T_K$, at intermediate to strong band interaction.
While there is agreement on the fact that a small $U_B$ enhances the effective Kondo 
coupling\cite{Khaliullin 95,Tornow 97,Davidovich 98,Takayama 98} 
but still leads to an exponentially vanishing $T_K$ at small hybridization
$\Delta$, it was found in ref.~\cite{Davidovich 98} that above an intermediate 
$U_B$ the Kondo temperature varies linearly in $\Delta$.
We will now consider this issue in detail.

We define $T_K$ to be equal to the binding energy of the local 
singlet, which is given by the position of the maximum in 
$\chi''(\omega)$. Note that this definition also applies to
the case of a noninteracting impurity (U=0).
With increasing $\Delta$ we observe a crossover 
from an exponential to a power law behavior $T_K \sim \Delta$.
The crossover point depends on $U_B$ and is proportional 
to the effective bandwidth $D_{\rm eff}$. 
For very small $\Delta$, the Kondo temperature always varies as
$\ln T_K\sim - U/\Delta$.
In contrast to ref.~\cite{Davidovich 98} we therefore 
find an exponentially small $T_K$ at any $U_B$, as 
long as the host is metallic. 
The discrepancy may be due to the approximate variational 
method used in ref.~\cite{Davidovich 98}.

\begin{figure}[t]
\begin{center}
\epsfig{file=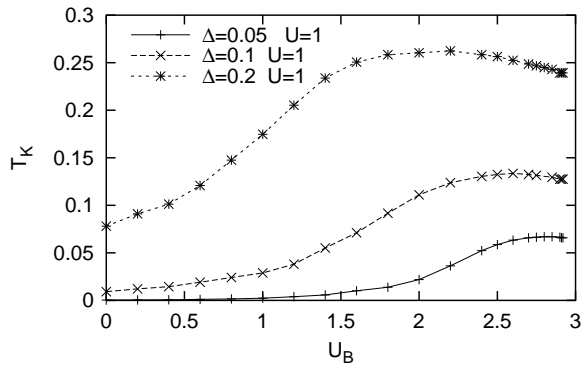,width=0.9\linewidth}
\end{center}
\caption{\label{fig:TK_3}Kondo temperature as a function of the 
band interaction.}
\end{figure}

At a finite band interaction, $U_B$ can lead to a
non-monotonic behavior of $T_K$; see fig.~\ref{fig:TK_3}.
The increase at small $U_B$ can be attributed 
to the local interaction on the site $i=0$ while the decrease 
close to the MIT is due to band narrowing. As $U_B\to U_{\rm MIT}$, 
the Kondo scale approaches a finite limiting value, indicating 
that even in the paramagnetic insulator the local impurity 
is screened! We can understand this by considering the 
effective hybridization ``seen'' by the $f$-impurity
\beq
\Delta_f = \frac{V^2}{\omega + i 0^+ - \Delta_c(\omega^+)}.
\eeq
In the insulating host $\Delta_c = 0$ and therefore 
\beq
\Delta_f(\omega) \sim V^2\, \delta(\omega).
\eeq
In this case, the impurity couples exclusively to the 
\mbox{$i=0$} site, the singlet is purely local and no 
Kondo many-particle physics is possible.
In conclusion, we analyzed a model describing an 
Anderson impurity in a correlated band. 
The limit of large coordination numbers made a treatment 
within the dynamical mean-field-theory possible. Using 
the nonperturbative numerical renormalization group, we solved the 
corresponding effective two-impurity model and obtained 
the one particle spectra and the dynamic 
susceptibility. 
We found that the system is always a Fermi liquid 
as long as the host is metallic.
Band correlations lead to a strongly enhanced Kondo scale, 
indicating that the dominant effect of $U_B$ is to increase the spin
polarization of the conduction electrons. 
Nevertheless, $T_K$ remains 
exponentially small as a function of hybridization. 
This is consistent with a Fermi liquid picture of the Hubbard host
where the Kondo screening of the impurity is due to fermionic quasiparticles 
instead of bare electrons.
In the spectral quantities, a change of the lineshape 
and the formation of side peaks is observed close 
to the Mott transition. This is explained by a narrowing 
of the effective conduction band.

%Concerning experiment, the system most frequently 
%cited\cite{Fulde 93} 
%in this context is the electron doped cuprate $Nd_{2-x} Ce_x Cu O_4$
%which displays\cite{Brugger 93} heavy fermion behaviour with a coherence 
%temperature $T^*\approx 0.3K$. From specific heat measurements 
%on the insulating undoped substance, one estimates an 
%exchange coupling $\alpha \approx 10^{-3}eV$ between the $Nd$ 
%spins and the neighbouring $Cu$ sites. The standard Kondo 
%estimate $T^* = D\, e^{-1/(2 N(0)\alpha)}$ leads to a 
%coherence temperature orders of magnitude below experimental 
%results. Although, strictly speaking, one is dealing with 
%a concentrated system (a periodic Anderson model would 
%be more appropriate here\cite{Schork 97})  this is considered as an example 
%for the influence of band correlations on single impurity 
%physics.

Of course it would be desirable to compare our results 
with experiments on systems that can actually be considered 
as dilute. One possibility might be to perform ESR measurements on 
rare earth systems, where impurities can be introduced into 
a correlated host in a controlled way.
Here the main experimental signal 
(the absorption $\chi''(\omega)$)
could be directly related to our model calculations. 

In future calculations we will study the effects of finite 
temperature, different fillings and antiferromagnetic 
order on our findings. 
We will also extend our analysis to 
thermodynamic and transport properties.

The authors would like to thank T.A.~Costi, H.-A.~Krug von Nidda, 
A.~Schiller and S.~Tornow for useful discussions.


\begin{thebibliography}{99}

\bibitem{Anderson 61}P.W.~Anderson, Phys.~Rev.~B {\bf 124}, 41
(1961).
\bibitem{Hewson 93}A.C.~Hewson, \emph{The Kondo Problem to Heavy Fermions}, 
Cambridge Studies in Magnetism Vol.\ 2 (Cambridge University Press, Cambridge 1993).
\bibitem{Wilson 75}K.G.~Wilson, Rev.~Mod.~Phys.~{\bf 47}, 773 (1975).

\bibitem{Krishnamurthy 80} H.R.~Krishna-Murthy, J.W.~Wilkins and 
K.G.~Wilson, Phys.~Rev.~B {\bf 21}, 1003 (1980).
\bibitem{Wiegmann 81}P.B.~Wiegmann, Phys.~Lett.~A {\bf 31}, 163 (1981).
\bibitem{Nozieres 74}P.~Nozi\`eres, J.~Low Temp.~Phys.~{\bf 17}, 31 (1974).
\bibitem{Tsvelick 83}A.M.~Tsvelick and P.B.~Wiegmann, Adv. Phys. {\bf
32}, 453 (1983).
\bibitem{Brugger 93}T.~Brugger, T.~Schreiner, G.~Roth, 
P.~Adelmann, and G.~Czjzek, Phys.~Rev.~Lett.~{\bf 71}, 
2481 (1993); \\
P.~Fulde, V.~Zevin, and G.~Zwicknagl, Z.~Phys.~B {\bf 92}, 133 (1993).
\bibitem{Khaliullin 95}G.~Khaliullin and P.~Fulde, Phys.~Rev.~B {\bf 52}, 
9514 (1995).
\bibitem{Tornow 97}S.~Tornow, V.~Zevin, and G.~Zwicknagl,
cond-mat/9701137.
\bibitem{Davidovich 98}B.~Davidovich and V.~Zevin, 
Phys.~Rev.~B {\bf 57}, 7773 (1998).
\bibitem{Phillips 96}P.~Phillips and N.~Sandler, Phys.~Rev.~B {\bf 53}, 
R468 (1996).
\bibitem{Schiller 95}A.~Schiller and K.~Ingersent,
Europhys. Lett. {\bf 39}, 645 (1997).
%\bibitem{Bulla 98}R.~Bulla, A.C.~Hewson, and Th.~Pruschke, 
%J. Phys.: Cond. Mat. {\bf 10}, 8365 (1998).
\bibitem{Gutzwiller 63}M.C.~Gutzwiller, Phys.~Rev.~Lett.~{\bf 10}, 
159 (1963); \\
J.~Hubbard, Proc.~Roy.~Soc.~London A {\bf 276}, 238
(1963); \\
J.~Kanamori, Prog. Theor. Phys. {\bf 30}, 275 (1963).
\bibitem{Metzner 89}W.~Metzner and D.~Vollhardt, Phys.~Rev.~Lett.~{\bf 62}, 324 (1989).
\bibitem{Georges 96}A.~Georges, G.~Kotliar, W.~Krauth, and 
M.J.~Rozenberg, Rev.~Mod.~Phys.~{\bf 68}, 13 (1996).
\bibitem{Bulla 99b}R.~Bulla, Phys.~Rev.~Lett.~{\bf 83}, 136 (1999).
%\bibitem{Costi 99a}T.A.~Costi in \emph{Density-Matrix Renormalization - 
%A New Numerical Method in Physics}, Eds. I.~Peschel et al., Springer 
%1999.
\bibitem{Hofstetter 99}W.~Hofstetter and S.~Kehrein, Phys.~Rev.~B {\bf 59}, R12732 (1999).
\bibitem{Luttinger 61}J.M.~Luttinger, Phys.~Rev.~{\bf 121}, 942
(1961); \\
D.C.~Langreth, Phys.~Rev.~{\bf 150}, 516 (1966). 
\bibitem{Takayama 98}R.~Takayama and O.~Sakai, J.~Phys.~Soc.~Jpn.~{\bf 67}, 1844 (1998). 
\bibitem{Costi 99b}T.A.~Costi, private communication.
\bibitem{Shiba 75}H.~Shiba, Prog.~Theor.~Phys.~{\bf 54}, 967 (1975).
\bibitem{Yoshimori 82}A.~Yoshimori and A.~Zawadowski, 
J.~Phys.~C: Solid State Phys.~{\bf 15}, 5241 (1982).



\end{thebibliography}
\end{document}